\documentclass[10pt,a4paper,twoside]{article}
\usepackage{epsfig}
\usepackage{baltlat5}
\usepackage{wrapfig}
\begin{document}
\ \
\vspace{0.5mm}

\setcounter{page}{1}
\vspace{5mm}

\titlehead{Baltic Astronomy, vol.\ts 14, XXX--XXX, 2005.}

\titleb{NON-LTE METAL ABUNDANCES IN V652\,HER AND HD\,144941}

%\footnotetext{\footnotestyle  Some footnote}

\begin{authorl}
\authorb{N.~Przybilla}{1}
\authorb{M.~F.~Nieva}{1,2} 
\authorb{U.~Heber}{1} and
\authorb{C.S. Jeffery}{3}
\end{authorl}

\begin{addressl}
\addressb{1}{Dr.\,Remeis-Sternwarte Bamberg, Sternwartstr.\,7, D-96049 Bamberg, Germany}

\addressb{2}{Observat\'orio Nacional, R. Gal. Jos\'e Cristino 77, 20921-400,
S\~ao Crist\'ov\~ao, Rio de Janeiro, RJ, Brasil}

\addressb{3}{Armagh Observatory, College Hill, Armagh BT61 9DG, Northern Ireland}
\end{addressl}

%If there is one instutition only:
%\begin{authorl}
%\authorb{E.~G.~Mei\v stas}{} and
%\authorb{G.~Anyman}{}
%\end{authorl}
%
%\moveright-3mm
%\vbox{
%\begin{addressl}
%\addressb{}{Institute of Theoretical Physics and Astronomy,
%Go\v{s}tauto 12, Vilnius LT-2600, Lithuania}
%\end{addressl}
%}

\submitb{Received 2005 July 31}

\begin{abstract}
Two evolutionary scenarios are proposed for the formation of extreme helium
stars: a post-AGB star suffering from a late thermal pulse, or the merger of two white dwarfs.
An identification of the evolutionary channel for individual objects~has~to
rely on surface abundances.
%of the light elements H, He and CNO, while the stellar metallicity is constrained 
%from the abundances of the heavier elements. 
We present preliminary results from a non-LTE analysis of CNO,
Mg and S for two unique objects, V\,652\,Her and HD\,144941.
Non-LTE abundance corrections for these elements range from 
negligible values to $\sim$0.7\,dex. Non-LTE effects typically lead to
systematic shifts in the abundances relative to LTE and reduce the
uncertainties. % in the abundance determination.
\end{abstract}


\resthead{Non-LTE metal abundances in the EHes V652\,Her and
HD\,144941}{Przybilla, Nieva, Heber \& Jeffery}

\sectionb{1}{INTRODUCTION}

Extreme helium stars (EHes) are a rare class of low-mass H-deficient
objects with spectral characteristics similar to those of B-giants.
Most of the two dozen known EHes~could be explained by post-AGB evolution,
linking R\,Cr\,B stars to Wolf-Rayet type central stars of planetary nebulae,
see Heber~(1986) and Jeffery~(1996) for reviews. The two 
stars studied here, V652\,Her and HD\,144941, are unique among the 
class members because of surface gravities too high for post-AGB
evolution and atypical surface abundances. A merger of two He
white dwarfs was suggested for the evolutionary origin of V652\,Her (Saio \&
Jeffery~2000). The chemical composition puts important observational
constraints on evolutionary scenarios. However, all elemental abundance analyses of
EHes to date are based on the {\it assumption} of local thermodynamic
equilibrium (LTE),
therefore being subject to potential systematic uncertainties. We investigate here which
improvements can be expected from a state-of-the-art non-LTE abundance analysis.

\sectionb{2}{MODEL CALCULATIONS, OBSERVATIONS \& STELLAR PARAMETERS}

The model calculations are carried out in a hybrid non-LTE approach, see
Przybilla et al.~(2005) for details.
In brief, the atmospheric structure computations are performed using
{\sc Atlas12} (Kurucz~1996) for an appropriate chemical mixture.  

\clearpage

\vbox{
\centerline{\psfig{figure=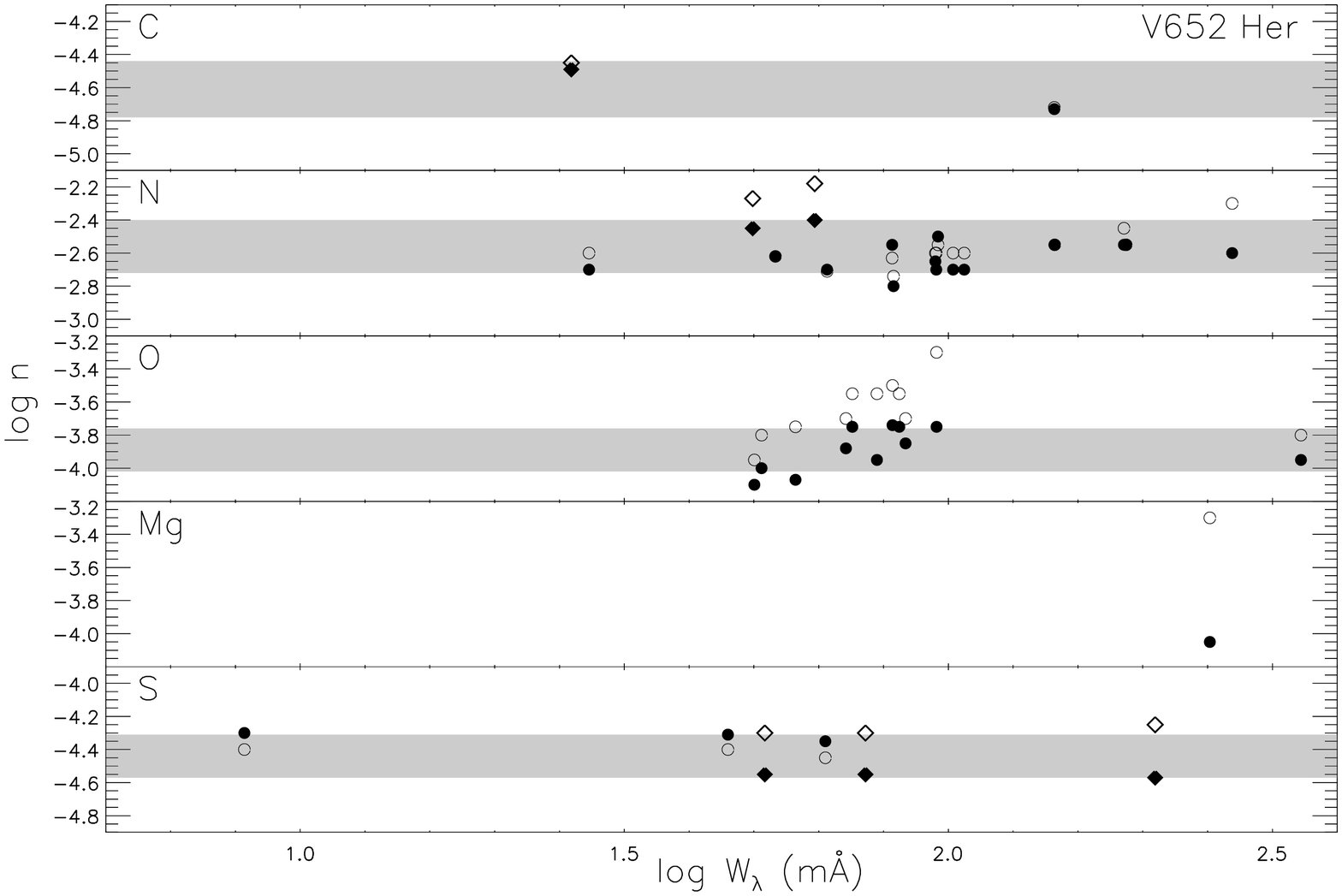,width=112truemm,clip=}}
\centerline{\psfig{figure=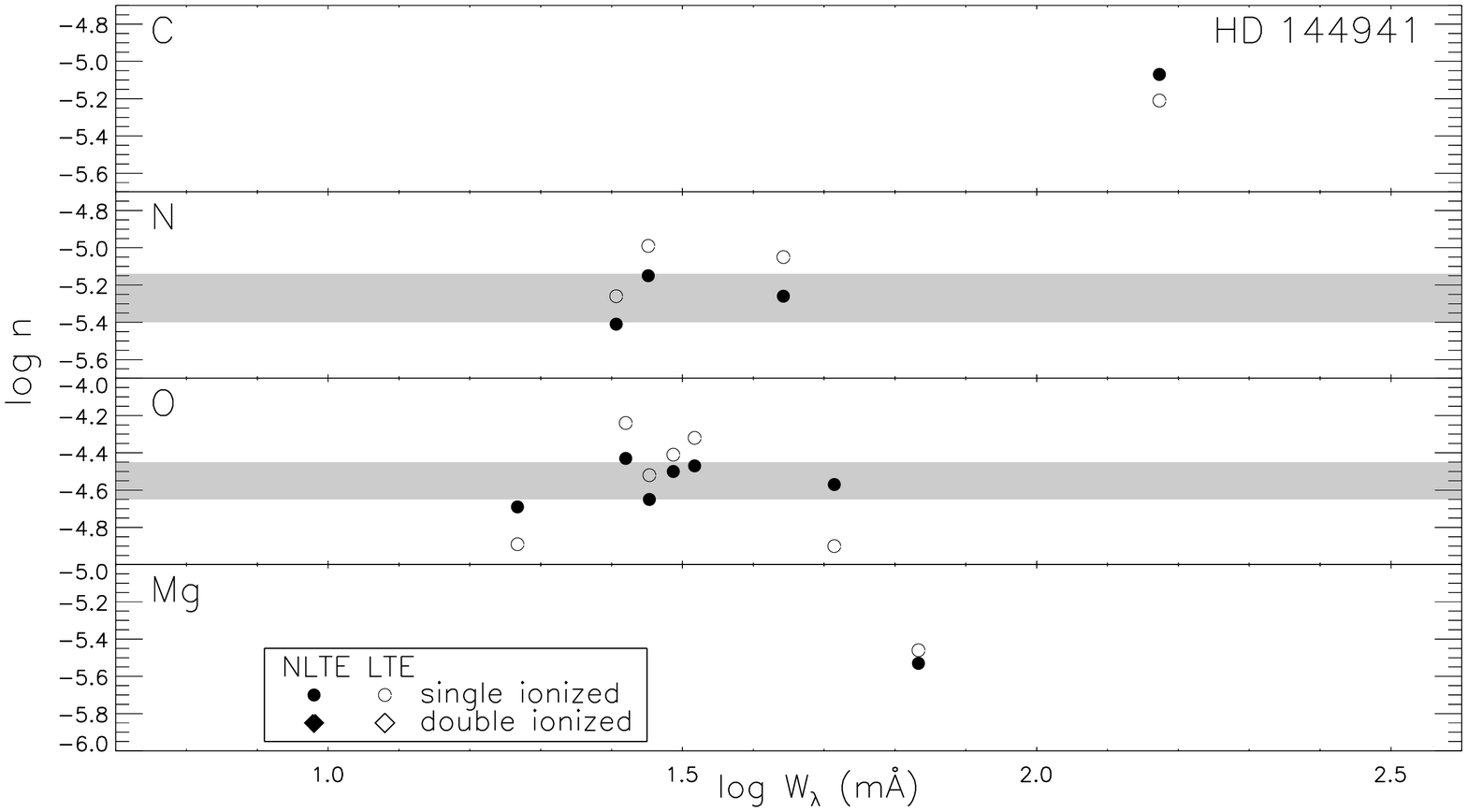,width=112truemm,clip=}}
\vspace{-2mm}
\captionc{1}{Elemental abundances in the sample EHes from individual spectral lines}
}
\vskip5mm

\parbox[b]{5.5cm}{\rule{-6mm}{0mm}
%\begin{wrapfigure}[12]{l}[0pt]{6.5cm}
%\vskip-2mm
%\vbox{
\tabcolsep=1.5pt
\begin{tabular}{ll}
\multicolumn{2}{c}{\parbox{5.3cm}{
~~~~{\bf Table 1.}{\ Non-LTE model atoms}}}\\
\tablerule
Ion & Source\\
\tablerule
H               & Przybilla \& Butler~(2004)\hhuad\\[-1pt]
He\,{\sc i/ii}  & Przybilla~(2005)\hhuad\\[-1pt]
C\,{\sc ii/iii} & Nieva \& Przybilla (in prep.)\hhuad\\[-1pt]
N\,{\sc ii/iii} & Przybilla \& Butler~(2001),\hhuad\\[-1pt]
                & with extensions\hhuad\\[-1pt]
O\,{\sc ii}     & Becker \& Butler~(1988)\hhuad\\[-1pt]
Mg\,{\sc ii}    & Przybilla et al.~(2001)\hhuad\\[-1pt]
S\,{\sc ii/iii} & Vrancken et al.~(1996),\hhuad\\[-1pt]
                & with updated atomic data\hhuad\\[-1pt]
\tablerule
\end{tabular}
%}
%\end{wrapfigure}
%\vskip2mm
}
\hfill
\parbox[b]{5.5cm}{\rule{-9mm}{0mm}
%\begin{wrapfigure}[9]{l}[0pt]{6.5cm}
%\vskip-2mm
%\vbox{
\tabcolsep=1.5pt
\begin{tabular}{lr@{$\pm$}lr@{$\pm$}l}
\multicolumn{5}{c}{\parbox{4.6cm}{
~~~~{\bf Table 2.}{\ Stellar parameters}}}\\
\tablerule
 & \multicolumn{2}{c}{V652\,Her\,($R_{\rm max}$)} &
\multicolumn{2}{c}{HD\,144941}\\
\tablerule
$T_{\rm eff}$\,(K)     & 22\,000 & 500               & 22\,000 & 1\,000\\
$\log g$               &    3.20 & 0.10              & 4.15    & 0.10\\
$\xi$\,(km/s)          &       4 & 1                 & 8       & 2\\
$n_{\rm H}^{\rm NLTE}$ &   0.005 & 0.0005            & 0.035   & 0.005\\
\tablerule
\rule{0cm}{1.66cm}
\end{tabular}
%}
%\end{wrapfigure}
}
\vskip2mm

\clearpage

\begin{wrapfigure}[29]{i}[0pt]{76mm}
\vskip-3mm
\psfig{figure=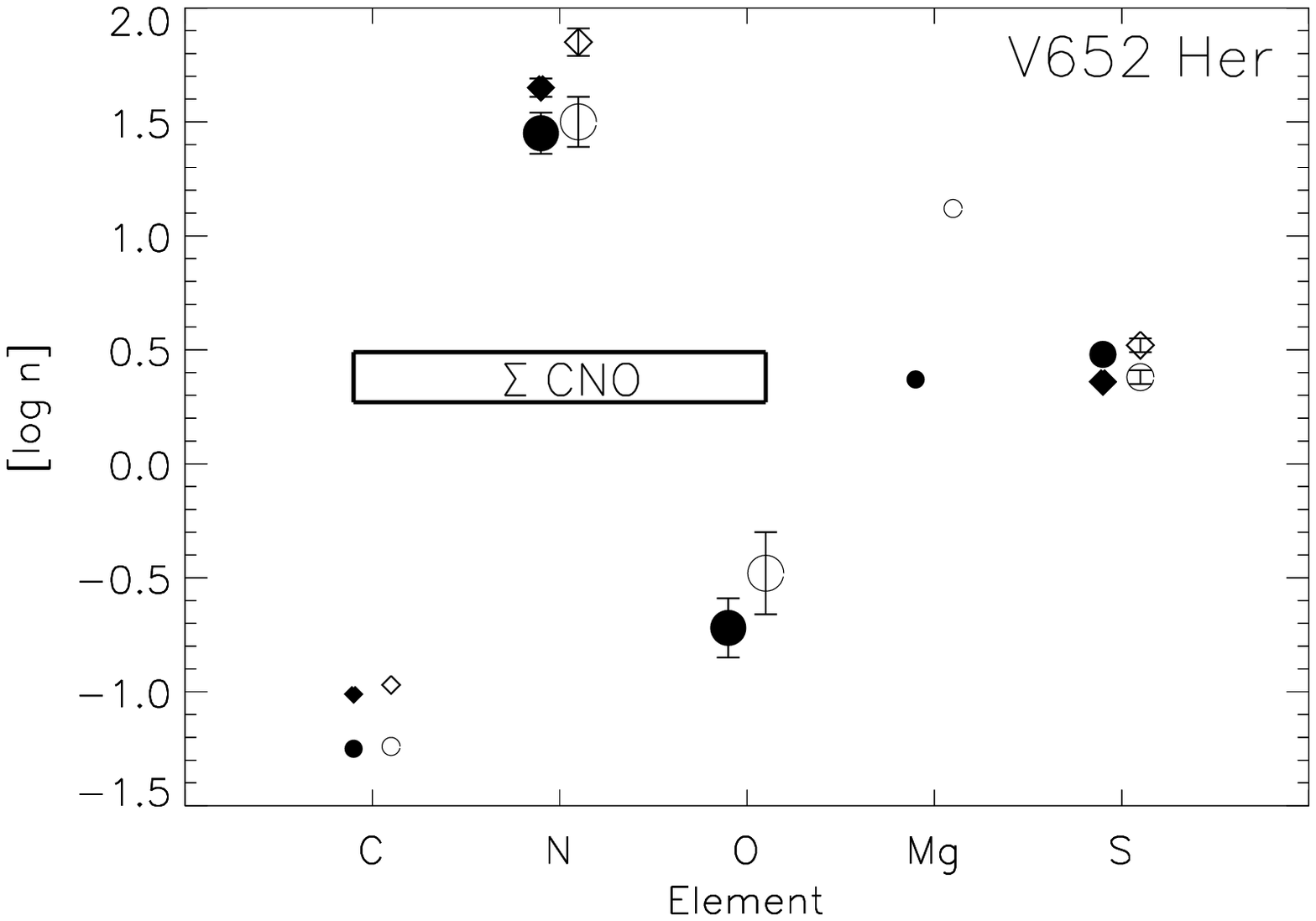,width=75truemm,clip=}\\[-5mm]
%\rule{0mm}{-4mm}
\psfig{figure=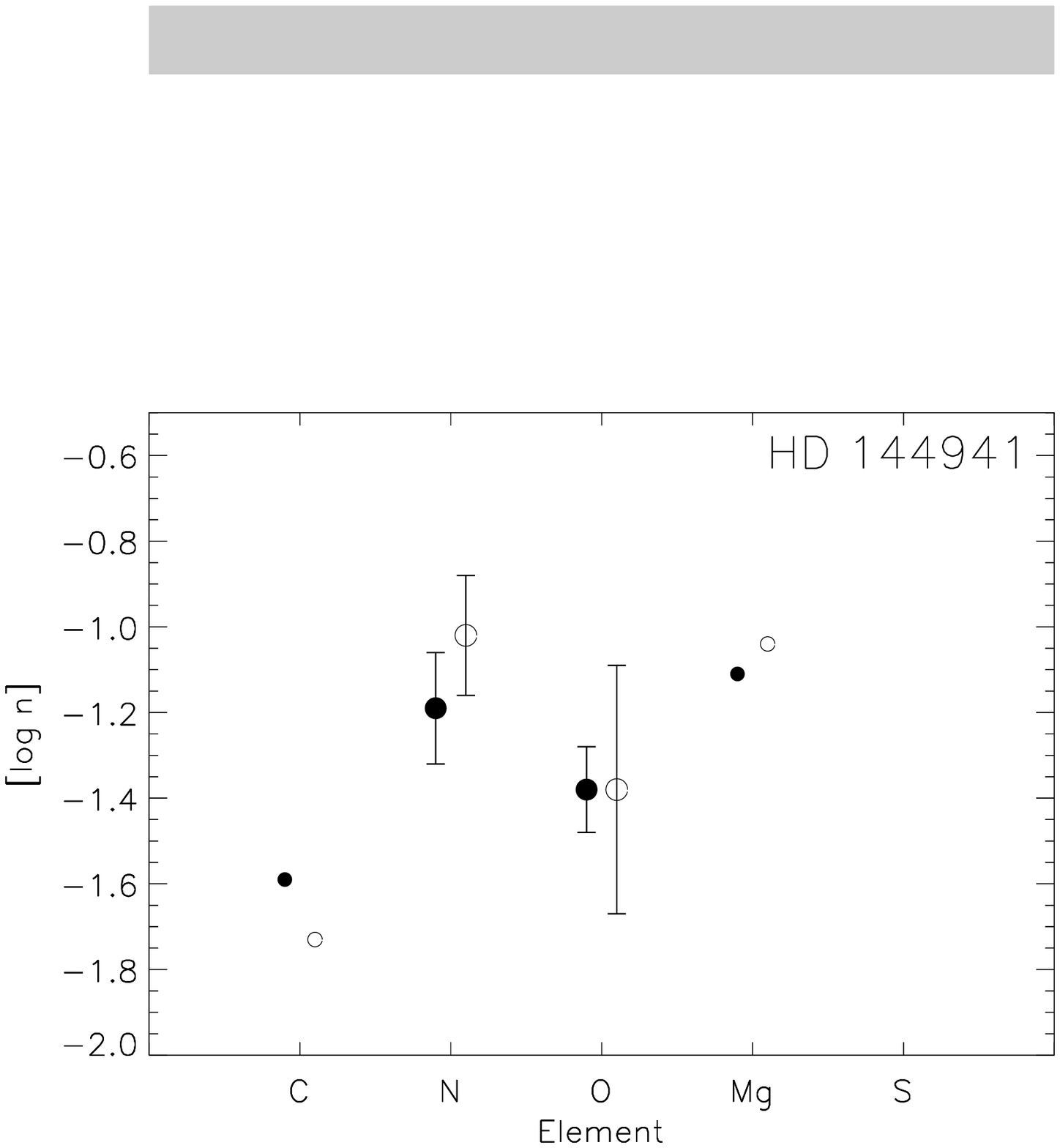,width=75truemm,clip=}\\[-8mm]
%\vskip-5mm
\captionc{2}{Metal abundances in the sample~EHes (symbols as in Figure~1)}
\end{wrapfigure}
\vskip5mm

\noindent Note that metal bound-free opacities are of minor
importance for the two objects.
Then, non-LTE line-formation is carried out on the resulting model stratifications.
Updated and improved versions of {\sc Detail} and {\sc Surface} 
(Giddings~1981; Butler \& Giddings~1985) are used for this, in combination
with state-of the-art non-LTE model
atoms, see Table~1. High-resolution spectra at high-S/N are adopted for the
analysis. Details on observations and data reduction are discussed
by Jeffery et al.~(2001) and Harrison \& Jeffery~(1997). The final model
parameters are summarised in Table~2, including microturbulence $\xi$ and
H abundance $n_{\rm H}^{\rm NLTE}$ (by number). They are derived from the
He\,{\sc i/ii} ionization equilibrium, Stark-broadened He\,{\sc i} lines and
H Balmer line strengths, using state-of the-art model atoms.
%for He and H (see also Table~1).
\vskip-2.5mm

\sectionb{3}{NON-LTE ABUNDANCES}

%Non-LTE abundances are determined only for C, N, O and the
%$\alpha$-elements Mg and S because we lack realistic non-LTE model atoms
%for the other chemical species. The lighter elements are of interest because
%of their involvement in fusion reactions, either as catalysts or as burning
%products, while the heavier elements can be used as tracers of 
%stellar metallicity. We intend to extend the study to Al, Si and Fe
%in the near future. Work on this is in progress.

Elemental abundances are derived from line-profile fits, using a 
$\chi^2$-minimisation technique based on small grids of synthetic spectra with 
varying metal abundances for given stellar parameters. This puts tighter constraints 
than the standard equi\-valent-width analysis. Abundances (by number) from 
individual spectral lines as a function of equivalent width $W_{\lambda}$ are displayed in 
Figure~1. 
%Non-LTE abundances are denoted by full and LTE results by open symbols;
%circles mark single-ionized and diamonds double-ionized species. 
The grey
bands indicate the 1$\sigma$-uncertainty range of the resulting abundances
for the chemical~species.

The non-LTE analysis reveals: {\sc i)} a reduction of systematic trends of
abundance with $W_{\lambda}$, {\sc ii)} systematic shifts in the abundances relative to
LTE, implying a downward revision in most cases and {\sc iii)} a potential for
reducing the statistical scatter by a significant amount, e.g. 
for O\,{\sc ii} in HD\,144941. Non-LTE abundance
corrections for individual lines can be as large as $\sim$0.7\,dex
(Mg\,{\sc ii} $\lambda$4481\,{\AA} in V652\,Her), but
usually they are (much) smaller. Note that the available spectra
cover only a restricted wavelength range, such that the analysis has to rely
on a rather small number of metal lines. Consequently, we view our results
as preliminary, also because of residual uncertainties in the 
stellar parameter determination, which are indicated by a slight mismatch in the metal
ionization equilibria in V652\,Her. The situation is more aggravated in
HD\,144941 because of its strong metal-deficiency that is larger than
in any other EHe star. The only metal ionization equilibrium available for
an independent verification of the stellar parameter determination is that
of silicon, for which we lack a reliable non-LTE model atom at present (the
model atom of Becker \& Butler~(1990) is too rudimentary in Si\,{\sc ii}).

The abundance patterns in the sample stars relative to the solar standard
(Grevesse \& Sauval~1998) are
discussed in Figure~2. For each ionic species non-LTE and LTE results with
uncertainties derived from the line-to-line scatter are displayed, using the
same symbols as in Figure~1. The symbol size encodes the number of lines
used for the abundance determination. 

Both stars exhibit CNO-processed material in their atmospheres. For V652\,Her 
the sum of CNO-abundances correlates well with the
super-solar metallicity as indicated by the magnesium and sulphur abundances. Note the
large non-LTE correction for magnesium, by a factor $\sim$5. Enhanced magnesium
abundances as derived from LTE analyses of several EHe stars have been controversial because
they cannot be consistently explained by nucleosynthesis (Jeffery~1996).
The current findings indicate a solution of this issue.
No further conclusions can be drawn for HD\,144941 at present. We intend to
extend the non-LTE study to aluminium, silicon and iron in the near future
in order to complete the diagnostic inventory.

The preliminary non-LTE analysis of the two unique hydrogen-deficient,
high-gravity objects V652\,Her and HD\,144941 does not drastically change our view of
their evolutionary origin as constrained from surface abundances. While the
non-LTE abundance corrections are small in most cases, they can be
highly important in other cases. Accounting for non-LTE effects improves on
the significance of abundance studies of extreme helium stars.

\References

\refb
Becker~S.~R., Butler~K. 1988, A\&A, 201, 232

\refb
Becker~S.~R., Butler~K. 1990, A\&A, 235, 326

\refb
Butler~K., Giddings~ J.~R.~1985, in Newsletter on Analysis of
Astronomical Spectra, No.\,9, Univ. London

\refb
Giddings~J.~R. 1981, Ph.\,D. thesis, Univ. London

\refb
Grevesse~N., Sauval~A.~J. 1998, Space Sci. Rev., 85, 161

\refb
Harrison~P.~M., Jeffery~C.~S. 1997, A\&A, 323, 177

\refb
Heber~U. 1986, in {\it Hydrogen Deficient Stars and Related Objects}, eds.
K.~Hunger, D.~Sch\"onberner \& N.~Kameswara Rao, Reidel
Publ. Co., Dordrecht, p. 33

\refb
Jeffery~C.~S. 1996, in {\it Hydrogen-Deficient Stars}, eds. C.~S.~Jeffery \&
U.~Heber, ASP Conf. Ser., 96, 152

\refb
Jeffery~C.~S., Woolf~V.~M., Pollacco~D.~L. 2001, A\&A, 376, 497 

\refb
Kurucz~R.~L. 1996, in {\it Model Atmospheres and Spectrum Synthesis}, eds.
S.~J.~Adelman, F.~Kupka, \& W.~W.~Weiss, ASP Conf. Ser., 108, 160

\refb
Przybilla~N. 2005, A\&A, 443, 293

\refb
Przybilla~N., Butler~K. 2001, A\&A, 379, 955

\refb
Przybilla~N., Butler~K. 2004, ApJ, 609, 1181

\refb
Przybilla~N., Butler~K., Becker~S.~R., Kudritzki~R.~P. 2001, A\&A, 369, 1009

\refb
Przybilla~N., Butler~K., Heber~U., Jeffery~C.S. 2005, A\&A, 443, L25

\refb
Saio H., Jeffery C.S. 2000, MNRAS, 313, 671

\refb
Vrancken~M., Butler~K., Becker~S.~R. 1996, A\&A, 311, 661

\end{document}